\documentclass[conference]{IEEEtran}
\usepackage{amsmath,graphicx,multirow,epsfig}
\usepackage[font=footnotesize]{caption}
\usepackage[font=footnotesize]{subcaption}
\usepackage{subcaption}
\usepackage{pgfplots}
\usepackage{pifont}
\usepackage{tikz}
\usepackage{array}
\usepackage{multicol}
\usepackage{graphicx}
\usepackage{algorithm}
\usepackage{algpseudocode}
\usepackage[bottom]{footmisc}

\ifCLASSINFOpdf
\else
\fi
\begin{document}
%
\title{Lossless Intra Coding in HEVC with Adaptive 3-tap Filters}

\author{\IEEEauthorblockN{Saeed Ranjbar Alvar and Fatih Kamisli}
\IEEEauthorblockA{Department of Electrical and Electronics Engineering, Middle East Technical University, Turkey \\
	 Eamils: Saeed.alvar@metu.edu.tr and Kamisli@eee.metu.edu.tr}
}


%


\maketitle
\begingroup\let\thefootnote\relax\footnotetext{This research was supported by Grant 113E516 of Tubitak}\endgroup

\begin{abstract}
In pixel-by-pixel spatial prediction methods for lossless intra coding, the prediction is obtained by a weighted sum of neighbouring pixels. The proposed prediction approach in this paper uses a weighted sum of three  neighbor pixels according to a two-dimensional correlation model. The weights are obtained after a three step optimization procedure. The first two stages are offline procedures where the computed prediction weights are obtained offline from training sequences. The third stage is an online optimization procedure where the offline obtained prediction weights are further fine-tuned and adapted to each encoded block during encoding using a rate-distortion optimized method and the modification in this third stage is transmitted to the decoder as side information. The results of the simulations show average bit rate reductions of 12.02\% and 3.28\% over the default lossless intra coding in HEVC and the well-known Sample-based Angular Prediction (SAP) method, respectively. \end{abstract}


%
\IEEEpeerreviewmaketitle

\section{Introduction}
Video coding standards such as the state of art High Efficiency Video Coding (HEVC) \cite{HEVC} and widely used H.264/AVC \cite{Luthra264} support both lossy and lossless compression. In both lossy and lossless compression modes, prediction is performed in a block based approach and then the difference between the original block and the predicted block (residual block) is further processed depending on the mode of compression and the input configurations. 

In lossless coding,  transform and quantization are skipped and the prediction residual block is directly entropy coded. In the mentioned standards, since the residual is obtained from a block based prediction which cannot provide a sufficiently well prediction for pixels away from the prediction boundaries, the energy of the residual block is high. In order to decrease the energy of the residual block, two set of approaches are proposed. In the first set  \cite{sulivanDPCM,cross,Secondary,IRDPCM,ARDPCM}, the residual block is post processed such that its energy is lowered. In the second set of approaches \cite{SAP,AD_SAP,DC2,Templatebased,fatihsaeed}, the prediction is obtained  by using pixel-by-pixel spatial prediction instead of block-based prediction. The details of the spatial prediction methods based on pixel-by-pixel predictions will be discussed in the following section. 

In this paper, a pixel-by-pixel spatial prediction method which uses three neighbouring pixels, similar to the algorithm discussed in \cite{fatihsaeed}, is proposed. However,  while \cite{fatihsaeed} obtains prediction weights offline from training sequences, this paper uses an online method to adapt the prediction weights to the content of each encoded block during encoding, which can provide further coding gains.
   
This paper is organised as follows. In section \ref{Spatial Pixel-wise} the spatial prediction methods based on pixel-by-pixel prediction will be discussed. Section \ref{three-tap filters} discusses the details of intra lossless coding with 3-tap filters. In section \ref{RDopt 3-tap} the proposed algorithm is introduced. Next section describes the details of the implementations the analyzes the performance of the implemented algorithms. Finally, the paper is concluded in section \ref{conclusion}. 

\section{Spatial prediction methods based on pixel-by-pixel prediction}
\label{Spatial Pixel-wise}
\begin{figure}[]
 \begin{subfigure}{\columnwidth}
	\centering
	\centerline{\includegraphics[ scale=0.6, trim = 5.1cm 21.0cm 4.9cm 2.0cm, clip=true]{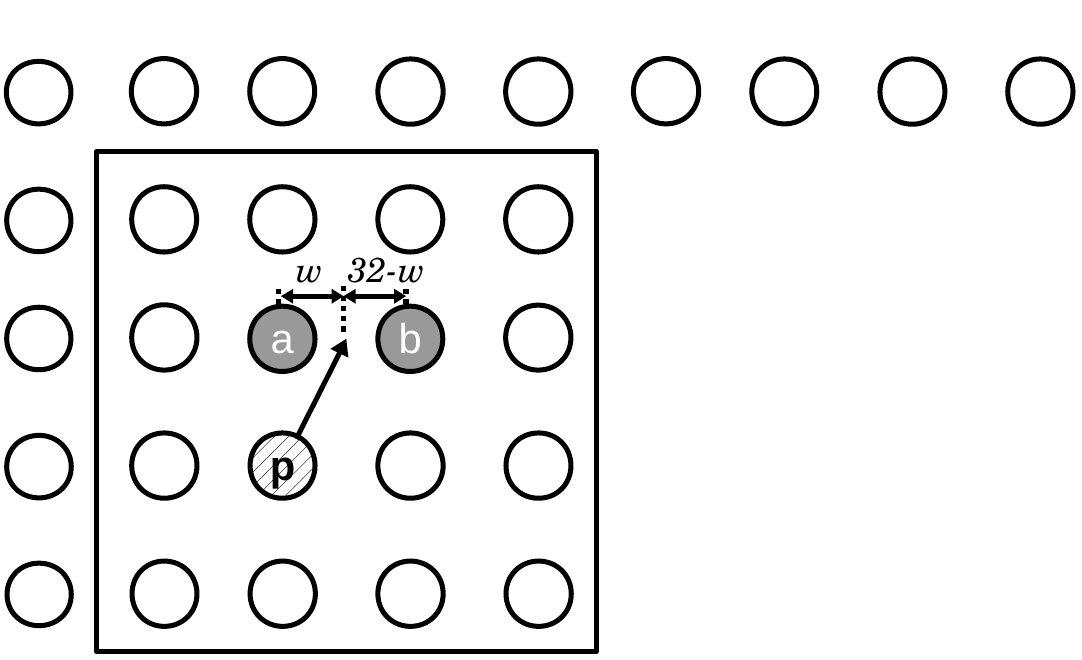}}
	\caption{Sample-based Angular Prediction (SAP) proposed for HEVC in \cite{SAP}. Block pixel to be predicted is projected on the immediately above row (or column, depending on mode) pixels (reference samples) along an angular direction. Prediction $p$ is obtained from linear interpolation of two closest reference samples $a$ and $b$.}
\end{subfigure}
 \begin{subfigure}{\columnwidth}	
	\centering
	\centerline{\includegraphics[scale=0.6,  trim = 5.1cm 21.0cm 4.9cm 2.0cm, clip=true]{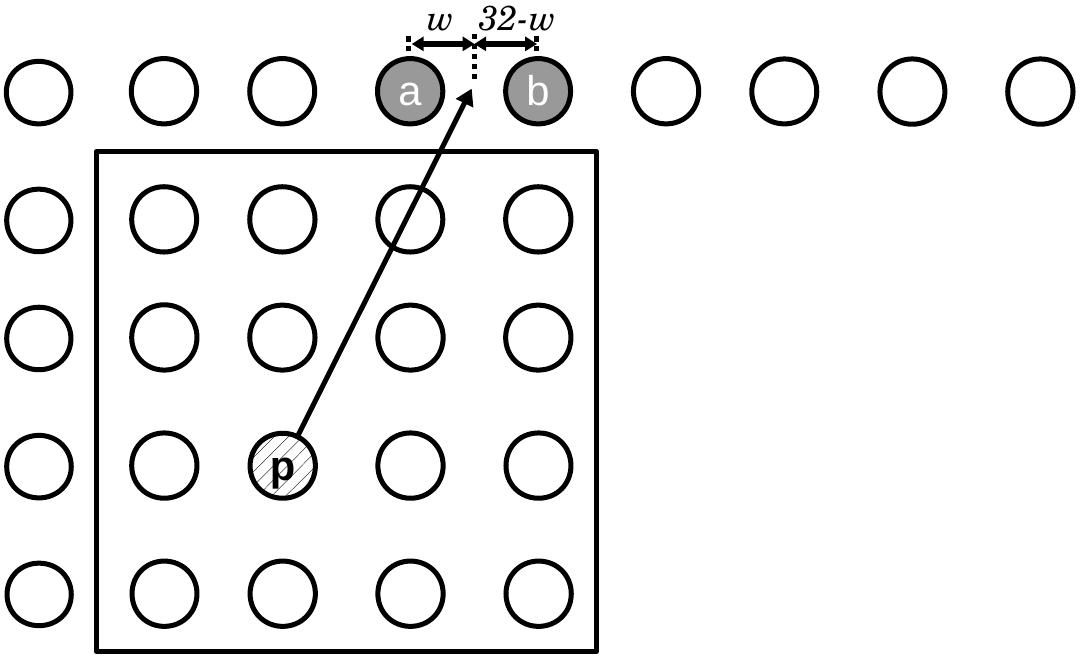}}
	\caption{Block-based angular prediction in HEVC. Block pixel to be predicted is projected on the block neighbor pixels (reference samples) along an angular direction. Prediction $p$ is obtained from linear interpolation of two closest reference samples $a$ and $b$.}
\end{subfigure}	
\caption{difference of the projection in SAP (a) and in HEVC (b)} 
\label{fig:sap}
\end{figure}

When the transform step is skipped in lossless intra coding, the block-based spatial prediction becomes less effective since some block pixels are predicted from distant reference samples and there is no transform step that can compensate for this inefficient prediction. However, since the transform is skipped in lossless coding, a pixel-by-pixel spatial prediction approach can now be used instead of a block-based approach for more efficient prediction.\cite{fatihsaeed}
Sample-based Angular Prediction (SAP) \cite{SAP} is an example of a pixel-by pixel prediction. In SAP, Modes Planar and DC are kept as HEVC's default modes and the angular modes are modified. In the directional modes the prediction angle and the formula utilized to form the output is the HEVC's angle and formula, however, there is one key difference between these two.  HEVC projects the current pixel into the reference samples (i.e. neighbor pixels of the block), on the other hand, SAP uses the same direction but the current pixel is projected to the immediate neighboring pixels. After the projection, the two closest pixels to the location of the projection are linearly interpolated and the value of the predicted pixel is obtained using the following formula. 
\begin{equation}
\label{eq:ang_pre}
p=((32-w) \cdot a + w \cdot b+16)>>5.
\end{equation}
Here "$>>$" indicates a bit shift operator, $a$ and $b$ represent the reference samples (closest pixels to the projected location), and $32-w$ and $w$ represent 5-bit integer interpolation weights, which are determined by the angle or prediction mode \cite{IntraHEVC}. Figure \ref{fig:sap} is an example of the projection of a pixel in SAP and HEVC.

Similar algorithm to SAP is adaptive directional SAP (AD-SAP)\cite{AD_SAP}. In this approach the prediction is similar SAP\cite{SAP} but encoder may change the direction of the prediction if it can remove the spatial redundancy more efficient than the ordinary SAP. 

One other example for algorithms of this type is Piecewise DC prediction where the average of the left and above pixels adjacent to the current pixel in the prediction unit (PU) is utilized for obtaining the prediction \cite{DC2}.

Another approach based on pixel by pixel prediction method is discussed in \cite{fatihsaeed}.  In this method three neighboring pixels are used for prediction according to a two-dimensional correlation model. The details of this algorithm is discussed in the following section. 

\section{Lossless Intra Coding with 3-tap Filters}
\label{three-tap filters}

\subsection{3-tap filtering approach}
In the most of the above mentioned algorithms, two neighboring pixels are used to obtain the prediction. The algorithm discussed in \cite{fatihsaeed} adds the third pixel and the weighted sum of these three pixels is the final value of the prediction for all intra modes.  In order to represent the directionality of intra modes in a more efficient way, these three neighbors are not fixed and they change depending on the mode of the prediction. In this method, the value of the prediction is obtained using the following equation:
\begin{equation}
\label{eq:3taps}
p = \rho_{1,k} \cdot a + \rho_{2,k} \cdot b + \rho_{3,k} \cdot c.
\end{equation} 

\begin{figure}[b]
 	\begin{minipage}[b]{0.21\linewidth}
 		\centering
 		\centerline{\includegraphics[scale=0.70, trim = 6.0cm 19.5cm 13.2cm 4.7cm, clip=true]{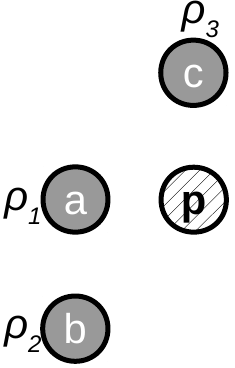}}
 		\begin{footnotesize}\centerline{(a) Modes 2-9}\end{footnotesize}\medskip
 	\end{minipage}
 	\begin{minipage}[b]{0.25\linewidth}
 		\centering
 		\centerline{\includegraphics[scale=0.70, trim = 6.0cm 19.5cm 13.2cm 4.7cm, clip=true]{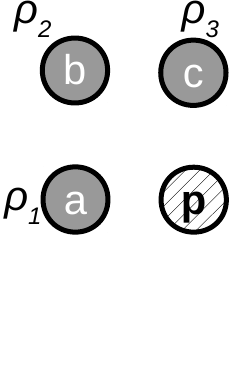}}
 		\begin{footnotesize}\centerline{(b) Modes 0,1,10-18}\end{footnotesize}\medskip
 	\end{minipage}
 	\begin{minipage}[b]{0.25\linewidth}
 		\centering
 		\centerline{\includegraphics[scale=0.70, trim = 6.0cm 19.5cm 13.2cm 4.7cm, clip=true]{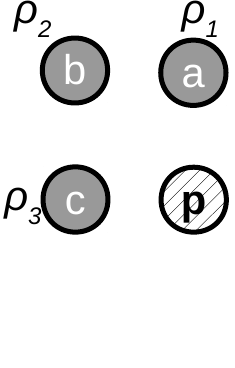}}
 		\begin{footnotesize}\centerline{(c) Modes 19-26}\end{footnotesize}\medskip
 	\end{minipage}
 	\begin{minipage}[b]{0.26\linewidth}
 		\centering
 		\centerline{\includegraphics[scale=0.70, trim = 6.0cm 19.5cm 11.9cm 4.7cm, clip=true]{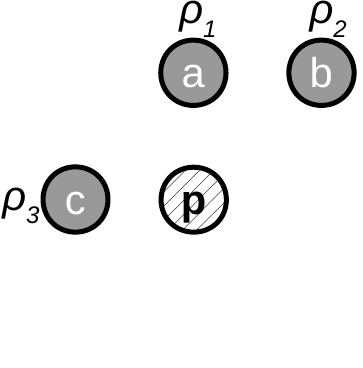}}
 		\begin{footnotesize}\centerline{(d) Modes 27-34}\end{footnotesize}\medskip
 	\end{minipage}
 	\caption{Neighbor pixels used for prediction in the 3-tap filtering method according to intra modes of HEVC. Intra modes 2-9 use neighbor pixels shown in (a), planar, DC modes (0,1) as well as intra modes 10-18 use neighbor pixels shown in (b), intra modes 19-26 use neighbor pixels shown in (c) and intra modes 27-34 use neighbor pixels shown in (d).}
 	\label{fig:3taps}
 \end{figure}

 $a$, $b$ and $c$ are the locations of neighboring pixels to be used in the prediction, see (Fig \ref{fig:3taps}) and $\rho_{1,k}$, $\rho_{2,k}$ and $\rho_{3,k}$ are weights corresponding to mode $k$.
 
\subsection{Prediction weights}
\label{Prediction weights}
The challenging issue with the prediction based on 3-tap filters is the value of the weights. The weights are obtained from a training sequence\footnote{A training sequence was formed from several images in the JPEG-XR image test set \cite{JPEGXR}.} using a two-stage optimization process. In the first stage, an iterative minimum squared-error(MSE) method is used, which finds the weights that minimize the mean squared prediction error over the training sequence. Let's $B_{opt}$ be the resulting bitrate from the parameters obtained from the first stage and  $\rho_{1,k}$, $\rho_{2,k}$ and $\rho_{3,k}$ be the prediction weights of intra mode $k$. Using $B_{opt}$ and  $\rho_{1,k}$, $\rho_{2,k}$ and $\rho_{3,k}$, the second stage of optimization is performed as following \cite{fatihsaeed}: 
\begin{enumerate}
	\item Let $k=0$ and $B_{best}=B_{opt}$.
	\item Generate 6 candidates for prediction weights of mode $k$ ($\rho_{1,k,i}$ , $\rho_{2,k,i}$ and $\rho_{3,k,i}$), run HEVC coder by replacing mode $k$'s weights with the candidates and record the resulting bitrates $B_i$.
	\begin{table}[h!]
		\caption{Candidate prediction weights}
		\label{tb:tesseq}
		\centering
		\begin{tabular}{c|c|c|c|c} 
			\hline   
			(Candidate) i & $\rho_{1,k,i}$   & $\rho_{2,k,i}$   & $\rho_{3,k,i}$   & Bitrate $B_i$ \\ 
			\hline \\[-2.0ex] \hline
			1 & $\rho_{1,k}+1$ & $\rho_{2,k}-1$ & $\rho_{3,k}$   & $B_1$  \\  \hline
			2 & $\rho_{1,k}-1$ & $\rho_{2,k}+1$ & $\rho_{3,k}$   & $B_2$  \\  \hline
			3 & $\rho_{1,k}+1$ & $\rho_{2,k}$   & $\rho_{3,k}-1$ & $B_3$  \\  \hline
			4 & $\rho_{1,k}-1$ & $\rho_{2,k}$   & $\rho_{3,k}+1$ & $B_4$  \\  \hline
			5 & $\rho_{1,k}$   & $\rho_{2,k}+1$ & $\rho_{3,k}-1$ & $B_5$  \\  \hline
			6 & $\rho_{1,k}$   & $\rho_{2,k}-1$ & $\rho_{3,k}+1$ & $B_6$  \\  \hline 
		\end{tabular}
	\end{table}
Find candidate with smallest bitrate, $i^{*}=argmin_{i} B_i$. If $B_{i^*}<B_{opt}$, update bitrate $B_{opt}=B_{i^*}$ and prediction weights for mode $k$,$\rho_{1,k}=\rho_{1,k,i^*}$ , $\rho_{2,k}=\rho_{2,k,i^*}$ and $\rho_{3,k}=\rho_{3,k,i^*}$.
	\item If $k<$number of modes, increment $k$ by one and go to step 2. If $k=$number of modes, check if this iteration over all intra modes improved bitrate, i.e. $B_{opt}<B_{best}$. If so, go to step 1, otherwise finish.
\end{enumerate}

While MSE and bitrate are in general coherent, reduction of MSE does not necessarily results in lower bitrate, therefore a second-stage optimization is performed to further fine-tune the weights to achieve minimum bitrate.
\section{LOSSLESS INTRA CODING WITH ADAPTIVE 3-TAP FILTERS}
\label{RDopt 3-tap}
According to the results discussed in  \cite{fatihsaeed}  the lossless intra coding using 3-tap filters provides substantial bitrate savings compared to HEVC lossless coding. In addition, it is shown that the second stage of the optimization discussed in \ref{Prediction weights} provides additional average 0.7\% bit rate reduction over the gains obtained from the first step. This proves that by further adjusting the weights of the filters we have the chance to obtain more accurate prediction. In other words, if we can modify the offline parameters of the 3-tap filters adaptive to the content of the current PU, we can reach a better prediction for that PU.   

As discussed earlier, the offline weights in \cite{fatihsaeed} are obtained from the data of a training set and those weights stay constant during the encoding and decoding procedure.  The algorithm that is proposed in this paper keeps the prediction as in \cite{fatihsaeed} but tries to use a technique similar to the second stage of the optimization discussed in \ref{Prediction weights} by modifying the $\rho$ parameters during the Rate Distortion Optimization (RDO) process. These adaptive weights help in finding the more accurate weights for the prediction. 

HEVC reference software, HM \cite{HM12},uses a three-step RDO method to find the optimal intra mode. In the first step, it returns N most promising modes out of 35 modes using Hadamard cost for RDO search. The N depends upon the PU size. The value of N can be \{8, 8, 3, 3, 3\} for PU size =\{4x4, 8x8, 16x16, 32x32, 64x64\}, respectively. Most Probable Modes (MPM) are then added to promising modes. In the second-step, RDO mode selection is performed and it returns the best mode based on rate distortion (RD) cost. Finally, the third step decides the best Transform Units (TU) partitioning for the current PU given the mode selected in the second-step\cite{HEVCRDO}. The first step is performed in order to avoid RDO mode selection to be tested for every possible intra mode. Hadamard Transform is chosen to provide more proper candidates for RDO mode selection process by simulating what transforms of HEVC do during RDO  but with a much simpler approach and with less number of operations. Since there is no transform in lossless case, using Hadamard cost is not efficient here. As a result, the first point that is suggested in this paper is to change the Hadamard cost to Sum of Absolute Difference between the original block and the prediction block (SAD cost) and find the candidates for RDO mode selection based on SAD cost.

The main goal which is further optimizing the $\rho$ parameters can be achieved in the first step of the described RDO process. The modified RDO process is as following: in the first step, similar to HEVC, N (stays as it is in HEVC) most promising modes out of 35 modes for a PU are obtained using the parameters in \cite{fatihsaeed} .Then before entering RDO mode selection step, for each of the N candidates, 8 set of $\rho$ parameters(1 as given in  \cite{fatihsaeed}, 6 as discussed in \ref{Prediction weights}, and 1 randomly chosen) are assigned and then sent to a new loop similar to first stage's loop which iterates 8 times. This 8 iterations, return 8 different SAD costs corresponding to each set of parameters. Then the best N among 8*N (N modes with 8 different set of parameters) cost that give the lowest SAD cost are proceeded to the next steps and finally the best mode and the corresponding set of parameters are resolved. It should be noted that we must signal which of those 8 sets achieved the lowest RD cost to be able to use it in other stages of the encoding and during the reconstruction in the decoder as well. Signaling the best candidate is done by a setting a 3 bit flag for each PU, which is written into the bit stream and is considered in cost calculation during the whole encoding process. Basically there are 7 modes for the flag to take (1  as given in \cite{fatihsaeed} and 6 as discussed in \ref{Prediction weights}). However, since a 3 bit flag is used , the eighth candidate is suggested randomly to use the capacity of 3 bits in signaling the candidates as much as possible.  Since 3 bits of redundant data is a considerable ratio of the  data size in  4x4 and 8x8 blocks, the improvement of the modified parameters cannot compensate  the cost of three bits in most of the cases. As a result, the proposed algorithm is applied for the remaining block sizes and  4x4 and 8x8 PUs are predicted using the offline parameters given in \cite{fatihsaeed} .Algorithm \ref{Alg1} summarizes the proposed algorithm.
\begin{algorithm}
	\caption{Proposed algorithm}\label{Alg1}
	\begin{algorithmic}[1]
		\State Start the RDO procedure
		\State Obtain the N most promising modes out of 35 using SAD cost		
		\For{each of N candidates }
		\State obtain additional 7 set of parameters and obtain the SAD cost for each.
		\EndFor
		\State Choose N candidates with the lowest SAD cost among 8*N candidates and update N most promising modes by setting the modes and the flags indicating which of 7 set of additional parameters exits in the list .
		\State Perform RDO mode selection and TU splitting stages and finalize the intra mode of PU.
		\State Write the information of the best mode and the corresponding flag into the bit stream.
	\end{algorithmic}
\end{algorithm}

\section{EXPERIMENTAL RESULTS} 
\label{exp}
The proposed prediction method based on adaptive 3-tap filters, method based on 3-tap filters using offline weights  \cite{fatihsaeed} and SAP \cite{SAP} are implemented in the HM12.0 \cite{HM12}. Initial 50 frames of the  sequences in Classes A to F are tested in  AI-Main configuration and QP=0, with the common test conditions \cite{commontest}. It should be noted that training sequences used in \ref{Prediction weights} don't include any of tested sequences. Table \ref{results} shows the average percentage bitrate reduction of three approaches compared to the lossless intra coding in HEVC.

From the comparison of the results, It can be seen that the proposed algorithm achieves the highest gains among the implemented algorithms. In addition, the comparison of the results of 3-tap filters with SAP shows how effective the third pixel is in removing the spatial redundancy. The results also reveal that the proposed algorithm has improved the gains of the 3-tap filters with the offline parameters for all the classes, especially for Class F where in average 0.96\% bit rate reduction is observed.

\begin{table}[]
	\centering
	\caption{Average percentage bitrate reduction of multiple methods and the adaptive 3-tap filtering method over HEVC lossless intra coding.}
	\label{results}
\begin{tabular}{|c|c|c|c|}
\hline
        & SAP\cite{SAP}   & \begin{tabular}[c]{@{}c@{}}3-tap filters\\  (offline weights) \cite{fatihsaeed}  \end{tabular} & \begin{tabular}[c]{@{}c@{}} adaptive 3-tap filters\\ \end{tabular} \\ \hline
        		\hline \\[-2.0ex] \hline
Class A & 8.94  & 15.60                                                    & 15.92                                                        \\ \hline
Class B & 5.78  & 8.59                                                     & 8.95                                                         \\ \hline
Class C & 6.95  & 8.67                                                     & 9.11                                                         \\ \hline
Class D & 8.91  & 11.06                                                    & 11.40                                                        \\ \hline
Class E & 10.56 & 14.35                                                    & 14.76                                                        \\ \hline
Class F & 12.51 & 12.48                                                    & 13.44                                                        \\ \hline
Average & 8.74  & 11.55                                                    & 12.02                                                        \\\hline
	\end{tabular}
\end{table}

\section{Conclusion}
\label{conclusion}
In this paper a novel pixel-by-pixel spatial prediction method based on the 3-tap filters is proposed for lossless intra coding of HEVC. In the proposed algorithm, despite the conventional prediction based on 3-tap filters,  the weights of the 3-tap filters are not constant and the modified value of the weights are explored adaptively during the RDO process.

The comparison of the performance of the proposed method with the HEVC's lossless gains in HM software, shows the average 12.02\% bit rate reduction. In addition, the proposed algorithm improves the performance of the intra prediction based on 3-tap filters with fixed weights, up to 0.96\% for some of the tested classes.


\section*{Acknowledgment}
This research was supported by Grant 113E516 of Tubitak.



\bibliographystyle{IEEEtran}
\bibliography{IEEEabrv,strings,refs}
%



\end{document}